\documentclass[a4paper,10pt]{article}
\usepackage[utf8]{inputenc}

\RequirePackage{hyperref}
\usepackage{amsthm,amsmath}
\usepackage{amssymb} 
\usepackage{authblk}
\usepackage{graphicx}

\emergencystretch=\maxdimen
\hypersetup{pdfauthor={Tommi Suvitaival, Simon Rogers, Samuel Kaski},
            pdftitle={Stronger findings from mass spectral data through multi-peak modeling},
            pdfkeywords={ANOVA-type modeling, Bayesian modeling, clustering, mass spectrometry, metabolomics, lipidomics, nonparametric Bayes}}

\title{Stronger findings from mass spectral data through multi-peak modeling}

\author[1]{Tommi Suvitaival\thanks{tommi.suvitaival@aalto.fi}}
\author[2]{Simon Rogers\thanks{simon.rogers@glasgow.ac.uk}}
\author[1,3]{Samuel Kaski\thanks{samuel.kaski@aalto.fi}}
\affil[1]{Helsinki Institute for Information Technology~HIIT, Department of Information and Computer Science, Aalto University}
\affil[2]{School of Computing Science, University of Glasgow}
\affil[3]{Helsinki Institute for Information Technology~HIIT, Department of Computer Science, University of Helsinki}

\begin{document}

\maketitle

\begin{abstract}

Mass spectrometry-based metabolomic analysis depends upon the~identification of spectral peaks by their mass and retention time. Statistical analysis that follows the~identification currently relies on one main peak of each compound. However, a~compound present in the~sample typically produces several spectral peaks due to its isotopic properties and the~ionization process of the~mass spectrometer device. In~this work, we investigate the~extent to which these additional peaks can be used to increase the~statistical strength of differential analysis.

We present a~Bayesian approach for integrating data of multiple detected peaks that come from one compound. We demonstrate the~approach through a~simulated experiment and validate it on ultra performance liquid chromatography-mass spectrometry~(UPLC-MS) experiments for metabolomics and lipidomics. Peaks that are likely to be associated with one compound can be clustered by the~similarity of their chromatographic shape. Changes of concentration between sample groups can be inferred more accurately when multiple peaks are available.

When the~sample-size is limited, the~proposed multi-peak\linebreak approach improves the~accuracy at inferring covariate effects. An~R implementation, data and the~supplementary material are available at~\href{http://research.ics.aalto.fi/mi/software/peakANOVA/}{http://research.ics.aalto.fi/mi/software/peakANOVA/}.

\end{abstract}

\section{Introduction}

The study of changes in the~levels of metabolites and lipids has become essential for the~comprehensive understanding of human health~\cite{Shevchenko10}. Mass spectrometry~(MS) techniques have become the~standard method for characterizing the~human metabolome~\cite{Scalbert09} and lipidome~\cite{Oresic08}. The~technique generates a~spectrum of peaks in the~plane defined by the~retention time and the~mass-to-charge ratio of an~observed particle. Each peak in this plane is either generated by a~particle arising from one of the~compounds present in the~sample, or is an~artifact of the~measurement without association to any of the~compounds. The~association between the~peaks and compounds is unknown \emph{a~priori}. The~produced peak data are noisy: First, sample preparation introduces sources of uncertainty that propagate to the~analysis~\cite{Dunn05}. Second, the~accuracy of the~device is limited~\cite{Windig96} and it produces biases. Third, peak identification, annotation and pre-processing steps produce 
additional layers of uncertainty~\cite{Smith06}. The~effect of errors at all these levels is exacerbated by the~``small~$n$, large~$p$'' problem: experiments cover a~very limited number of samples,~$n$, while the~set of compounds measured,~$p$, is potentially large.

\begin{figure}[h!]
 \centerline{\includegraphics[width=0.65\textwidth]{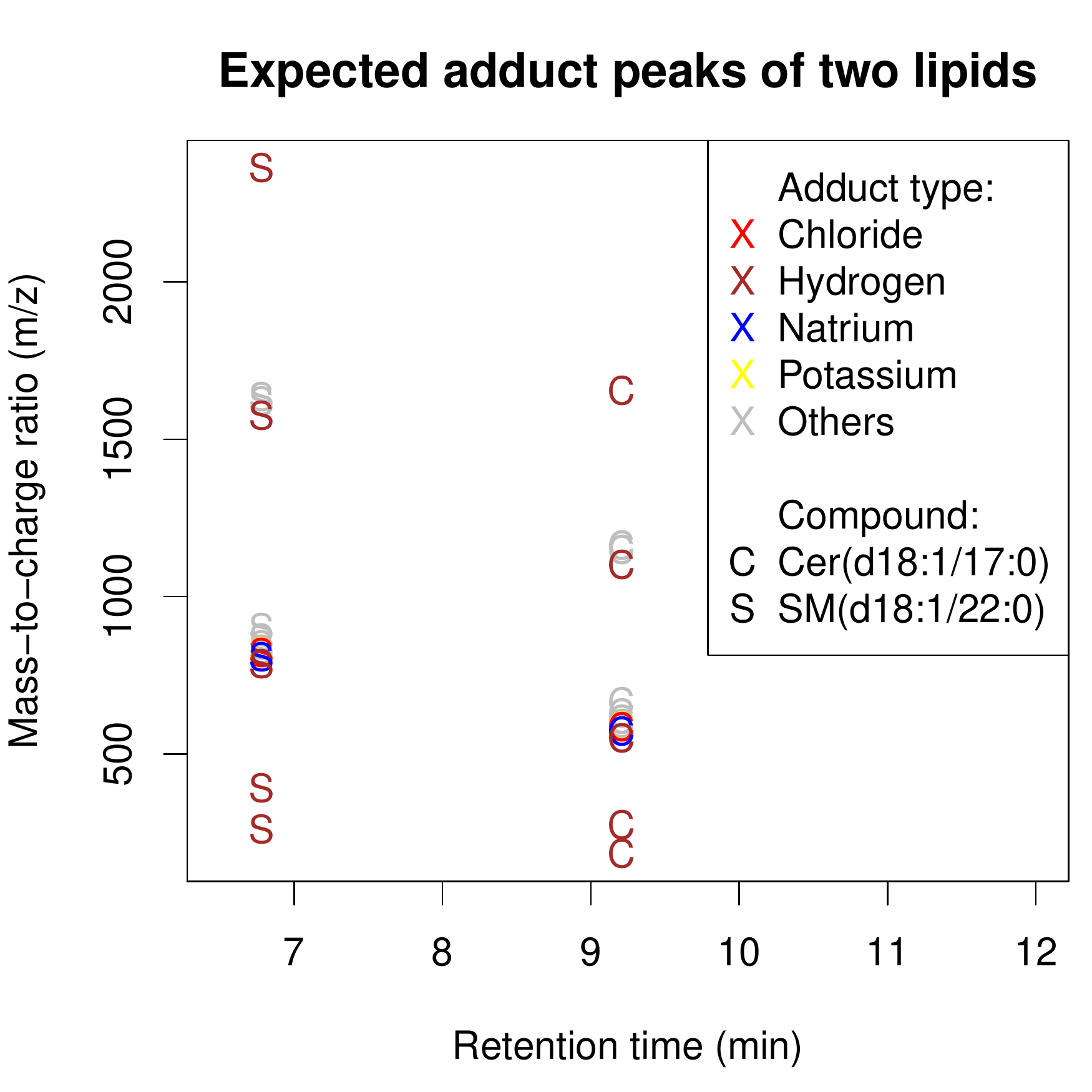}} 
 \caption{A schematic of the~positions of typical adduct peaks~\cite{Huang99} in the~RT-m/z plane for two lipids, the~ceramide~Cer(d18:1/17:0) and the~sphingomyelin~SM(d18:1/22:0). An~adduct peak is formed by an~ion attaching to the~compound. At~the~finer detail, each peak in the~figure consists of multiple isotope peaks few atomic units apart, as shown for~Cer(d18:1/17:0) in Figure~\ref{fig:isotopes}. Even though the~distinct isotope peaks are not visible to the~eye here, they are clearly separable by the~mass spectrometer. In~the~figure, adduct types and compounds are marked by colors and characters, respectively.}
 \label{fig:RTmzPlane}
\end{figure}

\begin{figure}[h!]
 \centerline{\includegraphics[width=0.65\textwidth]{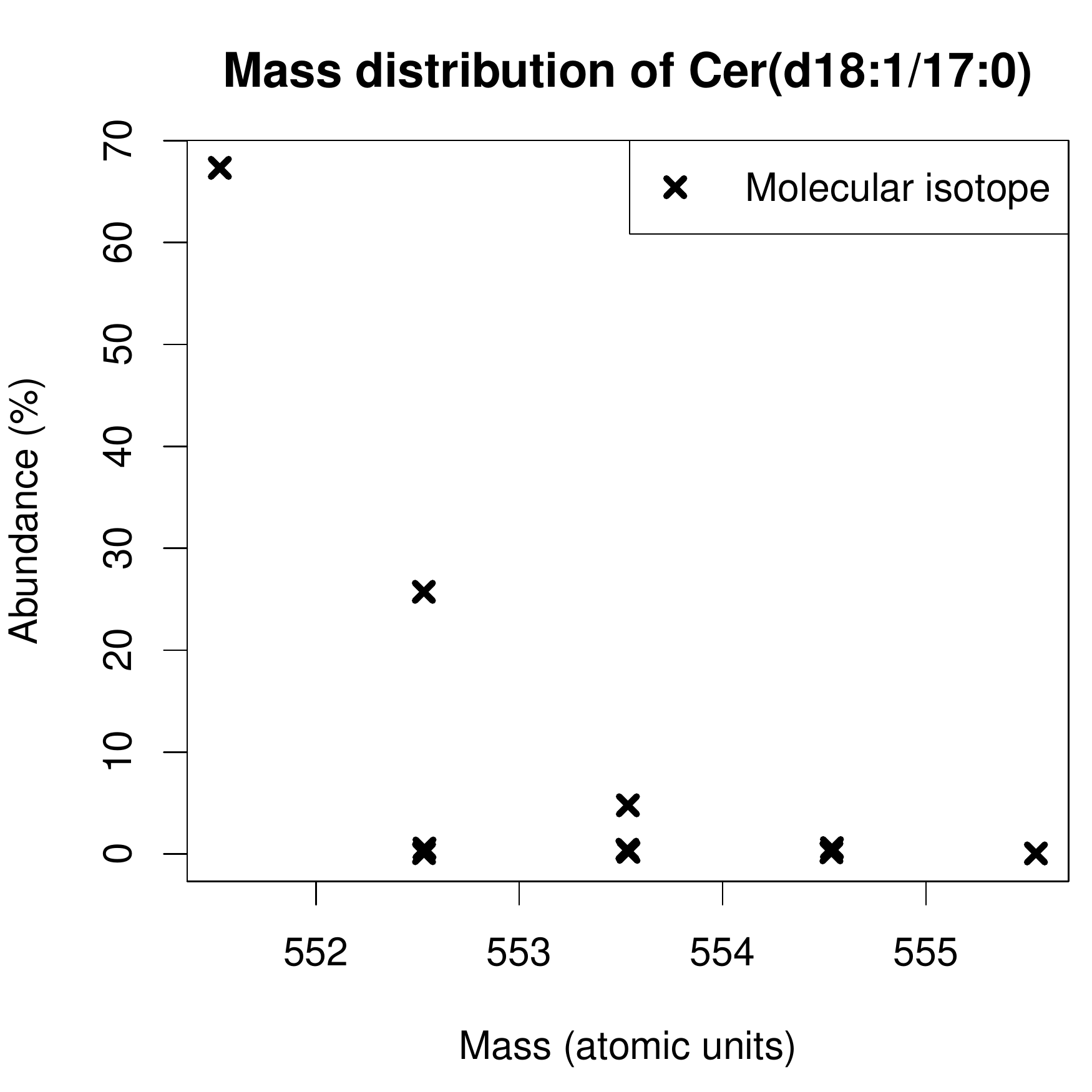}} 
 \caption{Natural isotopic distribution of the~mass of a~typical lipid, the~ceramide~Cer(d18:1/17:0). The~presence of atomic isotopes leads to the~appearance of multiple mass spectral peaks from the~compound. The~isotope peaks have distinct mass-to-charge ratios at the~same retention time (Figure~\ref{fig:RTmzPlane}).}
 \label{fig:isotopes}
\end{figure}

\begin{figure}[h!]
 \centerline{\includegraphics[width=0.65\textwidth]{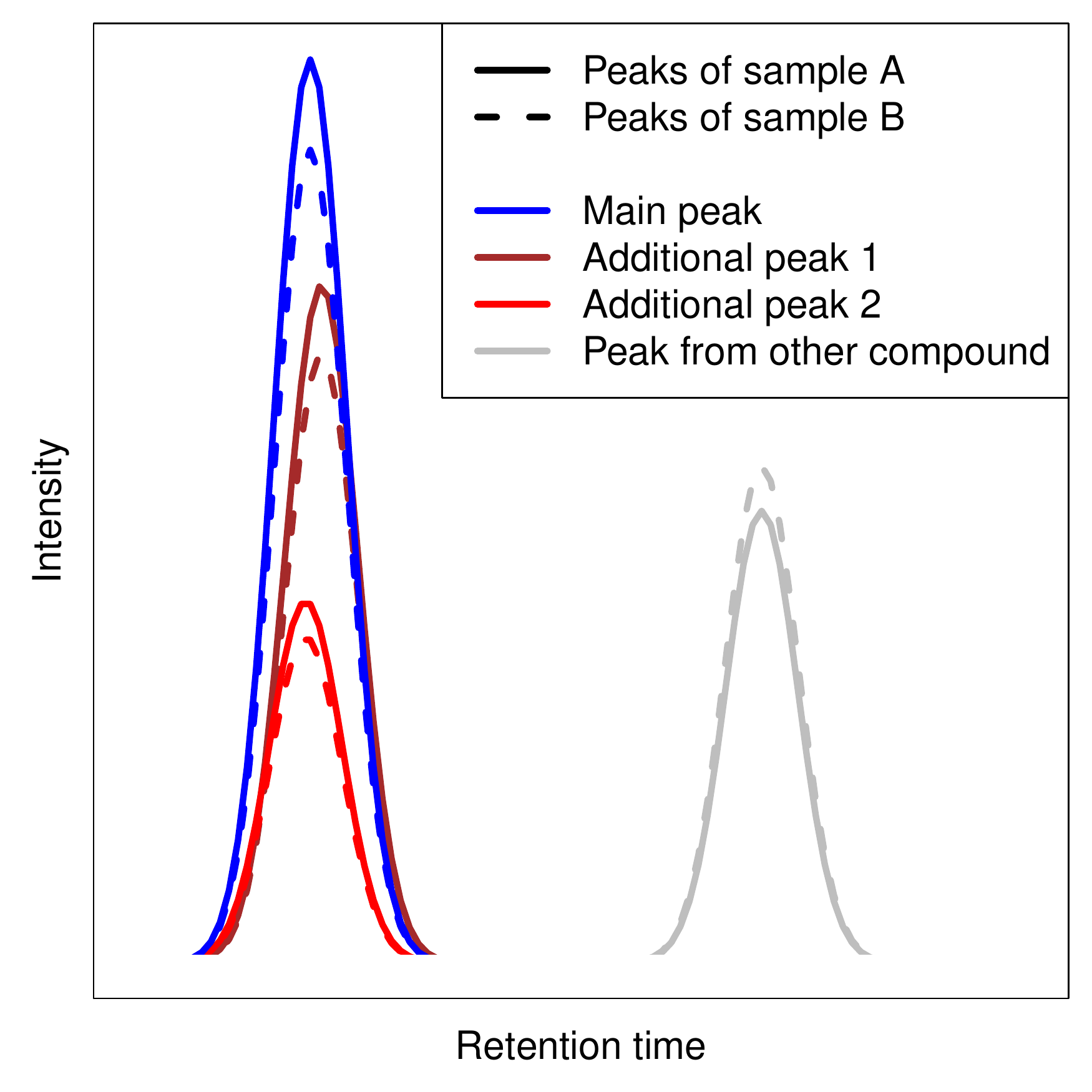}} 
 \caption{Schematic of spectral peaks in two samples. Peaks on the~left arise from one compound while peaks on the~right arise from another compound. The~heights of the~peaks from one compound have a~constant relative difference between the~two samples (solid and dashed lines), which can be used to improve the~accuracy of analysis. Distinct peaks from one compound~(marked by colors) have distinct mass-to-charge ratios but they share the~same retention time.}
 \label{fig:peaks-schematic}
\end{figure}

However, there also is strong informative structure in the~data: First, each compound may generate multiple adduct peaks~\cite{Huang99}~(Figure~\ref{fig:RTmzPlane}) and isotope peaks~\cite{Kind06,Bocker09} (Figure~\ref{fig:isotopes}), whose positions and shapes provide information about the~identity of the~compound. Second, the~concentrations of different compounds generated by or participating in similar biological processes may be highly correlated~\cite{Steuer06}. An~increasing number of machine learning algorithms are being developed for inferring such structure either from raw spectral data~\cite{Heinonen12} or from processed intensity data~\cite{Boccard10}. The~inference of covariate effects---the~differences between sample groups determined by external covariates such as an~intervention---is in the~core of the~comparative analysis of spectral profiles~\cite{Huopaniemi09}. In~this work, we propose an~unsupervised approach for inferring covariate effects from the~data more accurately through 
the~modeling 
of multiple peaks arising from a~single compound~(Figure~\ref{fig:peaks-schematic}).

The existence of additional peaks in the~spectrum is usually seen as a~problem and only the~main peak of each identified compound is used for further analysis. All peaks are a~result of the~ionisation process where a~charged particle is attached to or detached from a~compound. Each compound-ion pair produces a~distinct adduct or deduct peak. Random variation in the~ionisation process leads to inconsistencies between batches of samples, perceived as variation in the~ratio of intensities of the~peaks associated with one compound. This is a~major source of error for all existing analysis approaches regardless of the~choice of the~peak used for the~analysis. On~the~other hand, the~distribution of the~intensities of isotope peaks is by nature well preserved across both samples and compounds. Moreover, 
the~natural isotopic distribution of a~compound is known and can be used to make peak annotation more precise. In~this way, isotope peaks provide reliable additional information about the~differences in compound concentrations between sample groups.

We propose a~probabilistic approach for extending statistical analysis to all available peaks and show that the~additional peaks can provide a~real benefit to the~inference of covariate effects in an~experiment. The~approach is used to cluster the~peaks that are likely to arise from a~single compound together and to infer the~changes in concentrations of the~compounds more accurately based on all these peaks. By this approach, we are addressing the~problem of inadequate sample-size by introducing additional data describing the~compounds behind the~noisy measurements.

To solve the~problem we introduce the~following assumptions about the~generative process of the~data within a~Bayesian model: First, samples carry between-group differences in their compound concentrations and the~differences arise from responses to external covariates. Second, multiple observed spectral peaks follow an~identical generative process and their heights are a~noisy reflection of the~true concentration level of the~compound. Third, shapes of the~peaks from one compound are generated through an~identical process following the~properties of the~measurement device, and thus these shapes are highly similar.

The~approach presented in this paper consists of two stages of computational inference: (1)~peaks that share a~compound as their generative source are clustered together, and (2)~the~responses to external covariates are inferred on these groups of peaks.

The~clustering part of the~approach is based on a~nonparametric Bayesian Dirichlet process model~\cite{Rogers12}. To improve the~performance of the~clustering model, we have redefined the~prior distributions of the~model to have a~better match to the~distribution of the~peak shape similarity observations than in the~original proposition.

The~model for inferring the~responses to covariates operates on clusters inferred in the~first part. A~Bayesian multi-way model~\cite{Huopaniemi09} is suitable for this task. This model itself could be used for clustering summarized mass spectral intensity data, but in this work, we demonstrate that the~clustering can be done more accurately based upon the~similarity of chromatographic peak shapes.

\section{Methods}

This section describing the~models consists of two parts: clustering of spectral peaks and inference of covariate effects. To maintain the~mathematical rigor in the~section, we use the~terms ``samples,'' ``variables'' and ``clusters,'' to refer to the~experimental runs of the~mass spectrometer, the~peaks in the~mass spectrometry data, and the~yet unknown compounds in the~experimental runs, respectively. In~the~equations, we denote samples, variables and clusters by the~indices~$i$,~$j$ and~$k$, respectively:
\begin{align}
\begin{split}
i =& 1, \ldots, N \hspace{5pt} \text{(samples, \textit{i.e.}, experimental runs)},\\
j =& 1, \ldots, P \hspace{5pt} \text{(variables, \textit{i.e.}, peaks)},\\
k =& 1, \ldots, K \hspace{5pt} \text{(clusters, \textit{i.e.}, compounds)},
\end{split}
\end{align}
where~$N$,~$P$ and~$K$ are their total numbers.

\subsection{Clusters of peaks based on the~similarity}

Following earlier work~\cite{Rogers12}, we measure the~similarity between the~shapes of two peaks by their correlation computed over a~window of retention time after a~standard peak alignment~\cite{Pluskal10} across the~samples. Truncating negative values to zero, this leads to a~distinct similarity matrix~$\mathbf{Q}_{i \cdotp \cdotp}~\in~[0,1]^{P \times P}$ for each sample~$i$. In~the~notation, the~operator~``$\cdot$'' indicates that the~entire dimension of the~array is included, not only the~single item~$j$. Because a~peak is not necessarily observed in every sample, there may be missing values in the~matrices. Therefore, we construct an~additional mask~$\mathbf{R}~\in~\{0,1\}^{N \times P \times P}$ with binary values~$r_{ijj'}$ indicating whether peaks~$j$ and~$j'$ have any overlap in sample~$i$.

\subsubsection{Model}

We assume that the~peaks are generated through a~Dirichlet process~\cite{Escobar94}: there is an~unknown number of clusters and an~unknown and variable number of peaks that arise from each of the~clusters. Peaks are assumed to have a~one-out-of-many association: each peak is associated with only one of the~unknown clusters. With these basic assumptions, we can infer the~$P$-by-$K$ clustering matrix~$\mathbf{V}$ from the~data~$\mathbf{Q}$. To~make the~following equations more compact, we use an~additional variable~$\varepsilon_{jj'}~=~\mathbf{v}_{j\cdotp}\mathbf{v}_{j'\cdotp}^\text{T}~\in~\{0,1\}$, which is an~inner product of the~cluster indicator vectors of the~peaks~$j$ and~$j'$, to denote whether the~two peaks come from the~same or different clusters~(1~or~0, respectively).

We set a~spike-and-slab prior~\cite{Mitchell88} for the~peak shape similarity to model the~inherent sparse structure of the~data. The~similarity between any pair of observed peaks~($j,j'$) is assumed to follow a~beta distribution, but the~shape of the~distribution is assumed to depend on whether the~pair comes from the~same cluster or from different clusters~(shape parameters~$\left(a_\text{in},b_\text{in}\right)$ or~$\left(a_\text{out},b_\text{out}\right)$, when~$\varepsilon_{jj'}=1$~or~$0$, respectively). Also the~probability of a~missing similarity value is assumed to depend on the~cluster assignment of the~pair~($p_0^\text{in}$ or~$p_0^\text{out}$, when~$\varepsilon_{jj'}=1$~or~$0$, respectively). The~distributional assumptions are summarized~as
\begin{align}
p \left( q_{ijj'}, r_{ijj'} = 1 | \varepsilon_{jj'} = 1 \right) &= \left( 1-p_0^\text{in} \right) \text{Beta} \left( q_{ijj'} | a_\text{in}, b_\text{in} \right) \label{Eq:Cl-slabIn} \\
p \left( q_{ijj'}, r_{ijj'} = 0 | \varepsilon_{jj'} = 1 \right) &= p_0^\text{in} \delta \left( r_{ijj'} \right) \label{Eq:Cl-spikeIn}
\end{align}
for a~pair of peaks in the~same cluster~and
\begin{align}
p \left( q_{ijj'}, r_{ijj'} = 1 | \varepsilon_{jj'} = 0 \right) &= \left( 1-p_0^\text{out} \right) \text{Beta} \left( q_{ijj'} | a_\text{out}, b_\text{out} \right) \label{Eq:Cl-slabOut} \\
p \left( q_{ijj'}, r_{ijj'} = 0 | \varepsilon_{jj'} = 0 \right) &= p_0^\text{out} \delta \left( r_{ijj'} \right) \label{Eq:Cl-spikeOut}
\end{align}
for a~pair of peaks in different clusters. The~Dirac delta function~$\delta$, which is a~point mass at zero, constitutes the~``spike'' of the~two-component mixture distribution.

Combining Equations from~\ref{Eq:Cl-slabIn} to~\ref{Eq:Cl-spikeOut}, the~likelihood of the~entire peak shape data becomes a~product over all peak pairs and samples:
\begin{align}
\begin{split}
\mathcal{L} \left( \mathbf{Q}, \mathbf{R} | \mathbf{V} \right) = \prod_{i=1}^N \prod_{j=1}^{P-1} \prod_{j'=j+1}^P & p \left( q_{ijj'}, r_{ijj'} | \varepsilon_{jj'}=1 \right)^{\varepsilon_{jj'}} \\
& \times p \left( q_{ijj'}, r_{ijj'} | \varepsilon_{ijj'}=0 \right)^{1-\varepsilon_{jj'}} .
\end{split}
\end{align}

We introduce a~Dirichlet process prior for assigning peaks into clusters. In the~non-parametric model, the~probability of assigning peak~$j$ to an~existing cluster~$k$,
\begin{align}
 p \left( v_{jk} = 1 | \mathbf{Q}, \mathbf{R}, \mathbf{V}_{\cdotp,-k} \right) \propto s_k \mathcal{L} \left( \mathbf{Q}, \mathbf{R} | \mathbf{V}_{-j,\cdotp}, v_{jk}=1 \right) ,
\end{align}
is proportional to the~likelihood of the~peak given the~cluster, weighted by the~current size of the~cluster,~$s_k=\mathbf{v}_{-j,k}^\text{T}\mathbf{v}_{-j,k}$. In~the~notation, matrices~$\mathbf{V}_{\cdotp,-k}$ and $\mathbf{V}_{-j,\cdotp}$ correspond to the~matrix $\mathbf{V}$ with the~row~$k$ and the~column~$j$ omitted, respectively. Alternatively, with probability
\begin{align}
 p \left( v_{j,K+1} = 1 | \mathbf{Q}, \mathbf{R}, \mathbf{V} \right) \propto \alpha_\text{DP} \mathcal{L} \left( \mathbf{Q}, \mathbf{R} | \mathbf{V}_{-j,\cdotp}, v_{j,K+1}=1 \right) ,
\end{align}
the process may create a~new cluster with the~index~$K+1$ and only the~peak~$j$ inside. Now the~likelihood term is weighted by the~Dirichlet process concentration parameter~$\alpha_\text{DP}$.

\subsubsection{Inference}

We infer the~posterior distribution of the~clustering via Gibbs sampling, which results in a~set of~$S$ samples of the~clustering~$\mathbf{V}^{(s)}$,~$s=1,\ldots,S$, from the~true posterior distribution~$p \left( \mathbf{V} | \mathbf{Q}, \mathbf{R} \right)$. Further analysis can operate on the~entire posterior distribution of the~clustering through integration or on a~point estimate of the~distribution.

We follow earlier work~\cite{Dahl06} and acquire a~point estimate of the~posterior distribution of the~clustering through finding the~least-squares clustering: the~posterior sample~$s$ whose adjacency matrix~$\mathbf{V}^{(s)} \mathbf{V}^{(s) \text{T}}$ has the~smallest squared deviation from the~posterior pairwise probabilities~$\boldsymbol{\Pi}$ of the~peaks. The~index of the~least-squares clustering thus~is
\begin{equation}
 s_\text{LS} = \operatorname*{arg\,min}_{s \in \{1,\ldots,S\}} \sum_{j=1}^{P-1} \sum_{j'=j+1}^P \left( \mathbf{v}_{j\cdot}^{(s)} \mathbf{v}_{j'\cdot}^{(s) \text{T}} - \widehat{\pi}_{jj'} \right)^2 ,
\end{equation}
where the~posterior pairwise probabilities
\begin{equation}
 \widehat{\boldsymbol{\Pi}} = \frac{1}{S} \sum_{s=1}^S \mathbf{V}^{(s)} \mathbf{V}^{(s) \text{T}}
\end{equation}
are estimated as the~average adjacency matrix over the~posterior samples.

\subsection{Covariate effects based on peak heights}

Having inferred the~grouping of similar peaks into clusters that each correspond to a~compound, we infer the~differences in concentrations between sample groups for each cluster given the~peak height data~$\mathbf{X}~\in~\mathbb{R}^{P \times N}$ and the~clustering~$\mathbf{V}$. Again, there may be missing values in the~data.

\subsubsection{Model}

After a~peak-specific centering based on the~control group, the~observed peak heights for each sample~$i$ are assumed to be normally distributed with a~cluster-specific mean~$\mathbf{x}_{\cdotp i}^\text{lat}$:
\begin{equation}
 \mathbf{x}_{\cdotp i} | \mathbf{V}, \mathbf{x}^\text{lat}, \boldsymbol{\sigma}^2 \sim \mathcal{N} \left( \mathbf{V} \mathbf{x}^\text{lat}_{\cdotp i}, \boldsymbol{\Lambda} \right) ,
\end{equation}
where the~diagonal matrix~$\boldsymbol{\Lambda}$ contains the~peak-specific variance parameters~$\boldsymbol{\sigma}^2~\in~\mathbb{R}_+^P$. The~cluster-specific means are assumed to be normally distributed with a~sample group-specific prior~$\boldsymbol{\alpha}$, 
\begin{equation}
\label{eq:xlat}
 \mathbf{x}^\text{lat}_{\cdotp i} | \boldsymbol{\alpha}, a_i \sim \mathcal{N} \left( \boldsymbol{\alpha}_{\cdotp a_i}, \mathbf{I} \right) ,
\end{equation}
where~$a_i \in \{ 1, \ldots, L_a\}$ is an~indicator of group membership~(covariate level) for sample~$i$ and~$\mathbf{I}$ is a~$K\text{-by-}K$ identity matrix. The~corresponding covariate effects are arranged into an~$K$-by-$L_a$ matrix $\boldsymbol{\alpha}$ and the~effects are assumed to be independent and normally distributed,
\begin{equation}
\label{eq:alpha}
  \boldsymbol{\alpha}_{\cdotp l} \sim \left\{ \begin{array}{rl} \delta \left( \boldsymbol{\alpha}_{\cdotp l} \right), & l=1\\ \mathcal{N} \left( \mathbf{0}, \mathbf{I} \right), & l = 2, \ldots, L_a, \end{array} \right.
\end{equation}
except for the~first level,~$l=1$, which is defined as the~baseline level and thus is fixed to zero.

The model is not limited to one covariate. For an~arbitrary number of covariates, the~cluster-specific means~(Equation~\ref{eq:xlat}) can be expressed as a~sum of individual covariate effects and the~interaction effects of two or more covariates. This generalization can be written~as
\begin{equation}
 \mathbf{x}^\text{lat}_{\cdotp i} | \mathbf{E}, \mathbf{C} \sim \mathcal{N} \left( \mathbf{Ec}_{\cdotp i}, \mathbf{I} \right) ,
\end{equation}
where~$\mathbf{C}$ is an~$L$-by-$N$ binary matrix, whose column~$i$ indicates the~known levels of all the~covariates and all their resulting interaction levels for sample~$i$. The~matrix~$\mathbf{E}$ has dimensions~$K $-by-$L$ and its row~$k$ contains the~effects~$\mathbf{e}_{k\cdotp}$ of the~respective covariates and their interactions for cluster~$k$. Similarly, Equation~\ref{eq:alpha} can then be generalized~as
\begin{equation}
 \mathbf{e}_{\cdotp l} \sim \delta \left( \mathbf{e}_{\cdotp l} \right) ,
\end{equation}
when any of the~covariates is at the~base level,~and
\begin{equation}
 \mathbf{e}_{\cdotp l} \sim \mathcal{N} \left( \mathbf{0}, \mathbf{I} \right)
\end{equation}
otherwise.

The peak-specific variance parameter
\begin{equation}
 \sigma_j^2 \sim \text{Scale-Inv-}\mathcal{\chi}^2 \left( n_0, \sigma_0^2 \right)
\end{equation}
follows a~scaled inverse-$\chi^2$ distribution with~$n_0$ prior samples and a~scale~$\sigma_0^2$.

\subsubsection{Inference and analysis}

We infer the~covariate effects via Gibbs sampling. Now the~clustering matrix~$\mathbf{V}$ has been learned earlier, and is thus taken as known in the~model. The~posterior distributions of the~covariate effects~$\boldsymbol{\alpha}$ are descriptive of the~differences between the~sample groups and thus interesting from the~analysis point of view. To~assess the~significance of the~difference between a~sample group~$c>1$ and the~control group~$c=1$ for a~cluster~$k$, we can study the~posterior probability of the~effect~$\boldsymbol{\alpha}_{kc}$ being greater or smaller than zero.

\section{Results}

We demonstrate the~performance of the~proposed method through three experiments: a~simulated data experiment, a~spike-in benchmark experiment with known changes in concentrations, and a~gene silencing experiment with measurements of the~lipidome of cancer cells.

Evaluation of the~performance on real data sets is not a~trivial task, as there is no ground truth available: neither the~identity of the~peaks nor the~true effect sizes are known. Thus, we also use spike-in data, where the~true covariate effects are known, although only a~small number of the~peaks are annotated.

For the~simulated and benchmark experiments, we can compute the~mean squared error~(MSE) between inferred and true covariate effects and use that as an~evaluation metric. As~a~result of the~$\log$-transformation of the~intensity data, we are quantifying relative changes between sample groups, independent of the~average height of each peak. In~the~model, we thus assume that the~change is preserved across the~peaks of one compound, in relative terms. The~significance of the~difference in the~MSE of the~proposed approach and the~comparison method is tested by a~paired one-sided $t$-test. The~false discovery rate is controlled by the~Benjamini-Hochberg step-up procedure~\cite{Benjamini95} at level~0.01. Additionally for the~simulated experiment, we can study the~inference of the~statistical significance of effects, as the~true distribution of the~data is known.

To assess the~sensitivity of the~approaches to noise in natural lipidomic data lacking a~ground truth, we use two types of indirect evaluation: First, we study the~consistency of the~inferred covariate effects given a~prior assumption about their similarity. Second, we examine the~robustness of the~inferred covariate effects to noise. Finally, we demonstrate differences between the~multi-peak and single-peak approaches through examples of qualitative analysis of annotated peak clusters.

We compare the~performance of the~following approaches and refer to them as Models~1,~2 and~3:
\begin{enumerate}
 \item the~multi-peak approach using both peak shape and height information, as~proposed in this work~(nonparametric clustering of peaks by their shape similarity, inference of covariate effects on the~clusters based on the~height of the~peaks),
 \item the~multi-peak approach using peak height information only~\cite{Huopaniemi09}~(clustering of peaks and inference of covariate effects based on the~height of the~peaks only),
 \item the~typical single-peak approach~(inference of covariate effects by the~height of the~strongest annotated peak only).
\end{enumerate}
For the~studied real data sets, we discover that peak height information alone is not enough for clustering the~peaks into individual compounds due to the~high level of noise and strong correlations between compounds. Thus, for real data we compare Model~1 to Model~3 and highlight the~benefit gained by using peak shape information.

\subsection{Simulated data}

\subsubsection{Introduction}

We start by investigating the~performance of the~proposed approach on synthetic data, where the~true covariate effects are known. We focus on a~usual task in exploratory analysis of biological data: the~detection of significant non-zero covariate effects. We measure the~performance by the~accuracy at this task---the~ratio of true positive and true negative significant differences among all effects. We use the~95~\% posterior quantiles to determine the~significance. Additionally, we compute the~MSE to the~true effects and follow earlier work~\cite{Vinh10} at studying the~performance of the~two clustering models by computing the~normalized information distance~(NID) to the~true clustering.

The approaches are tested across a~set of potential experimental settings to study how the~observation of additional peaks and samples affect the~performance. Simulated data are generated by assuming the~latent structure of Model 1. The~following data parameters are varied on a~grid: sample-size~$N = 2 \times \{ 3, 7, 15 \}$ and peak-specific noise~$\sigma^2 = \{ 1, 5 \}$. Additionally, the~number of peaks per cluster are varied between~3,~7 and~15. Covariate effects~$\boldsymbol{\alpha}_{\cdotp 2} = [2,-1,0.5,0,0,0,0]$ are generated for each data set. The~experiment is repeated~100 times with independent data sets.

\subsubsection{Result} 
\begin{figure}[h!]
 \centerline{\includegraphics[width=\textwidth]{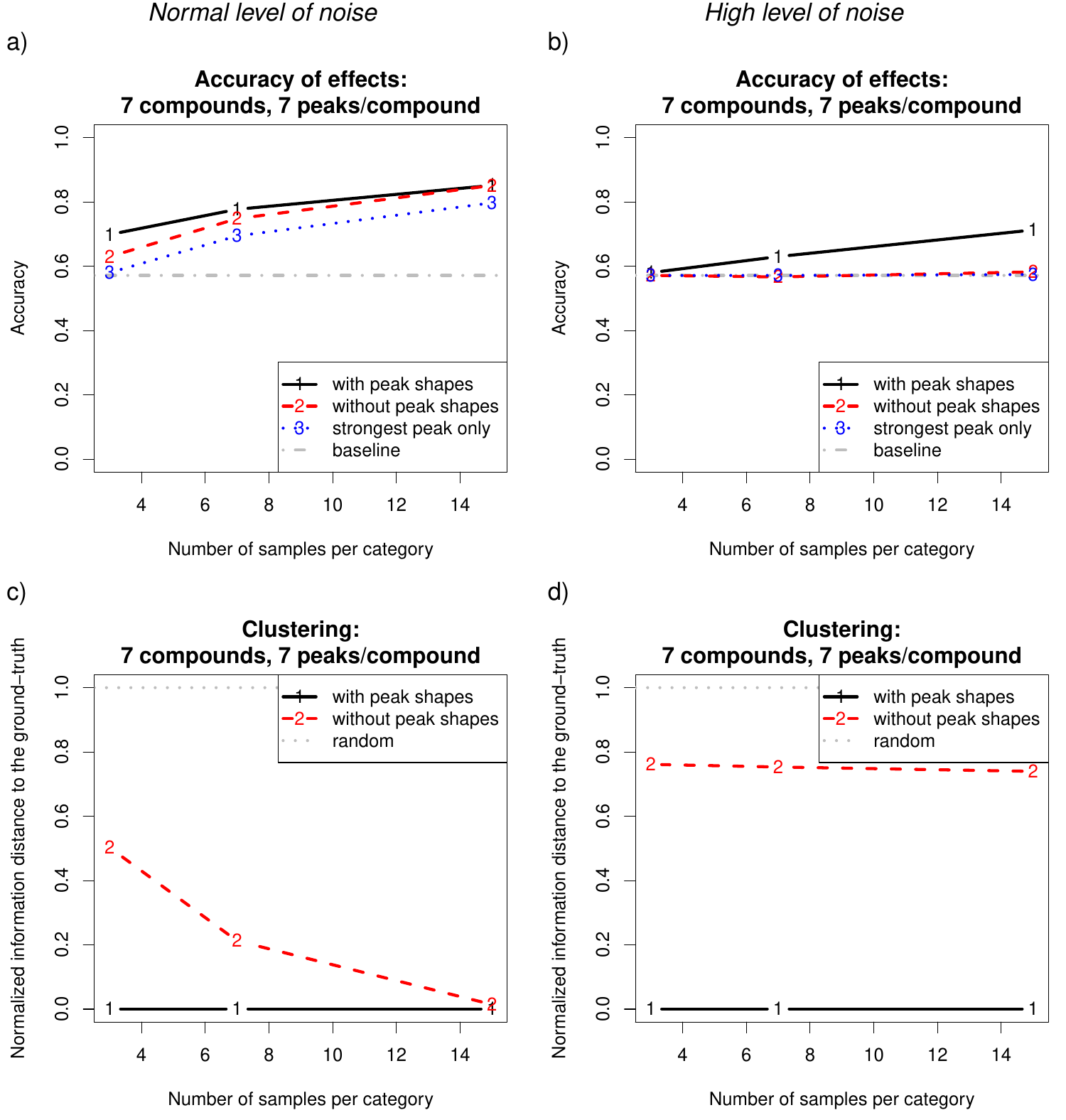}} 
 \caption{The use of data from multiple peaks and the~peak shape information increase the~accuracy at detecting significant covariate effects on simulated data. Accuracy of Models~1,~2 and~3 for simulated data is shown as a~function of the~sample-size in two settings: normal and high level of noise~(left:~$\sigma^2=1$, and right:~$\sigma^2=5$, respectively). Top~(a-b): Accuracy at inferring the~significance of the~generated covariate effects. Bottom~(c-d): Normalized information distance~(NID) between the~inferred and the~true clustering. An~entirely random and an~exactly correct clustering correspond to a~NID of~1 and~0, respectively.}
 \label{fig:simExp-Nsamples}
\end{figure}

On normal level of noise~($\sigma^2=1$), the~multi-peak approaches~(Models~1 and~2) always perform better at detecting significant covariate effects than the~single-peak approach~(Model~3; Figure~\ref{fig:simExp-Nsamples}a) and only with enough samples the~performance of Model~2 becomes comparable to Model~1. The~inferred clustering of Model~1 is perfect while the~clustering performance of Model~2 is heavily dependent on the~number of samples available~(Figure~\ref{fig:simExp-Nsamples}c).

On high level of noise~($\sigma^2=5$), only Model~1 works~(Figure~\ref{fig:simExp-Nsamples}b). The~reason for the~failure of Model~2 is the~imperfectly inferred clustering~(Figure~\ref{fig:simExp-Nsamples}d). The~good performance of Model~1 comes from the~clustering step, which is robust to noise in the~peak heights. The~peak shape similarity yields strong evidence for inferring the~clusters already from a~single sample.

The MSE between the~inferred and true covariate effects for Model~1 is smaller compared to Model~3 in all~24 setups of the~experimental grid. The~difference is statistically significant in~22 setups and in all setups at the~high level of noise.

\begin{figure}[h!]
 \centerline{\includegraphics[width=0.65\textwidth]{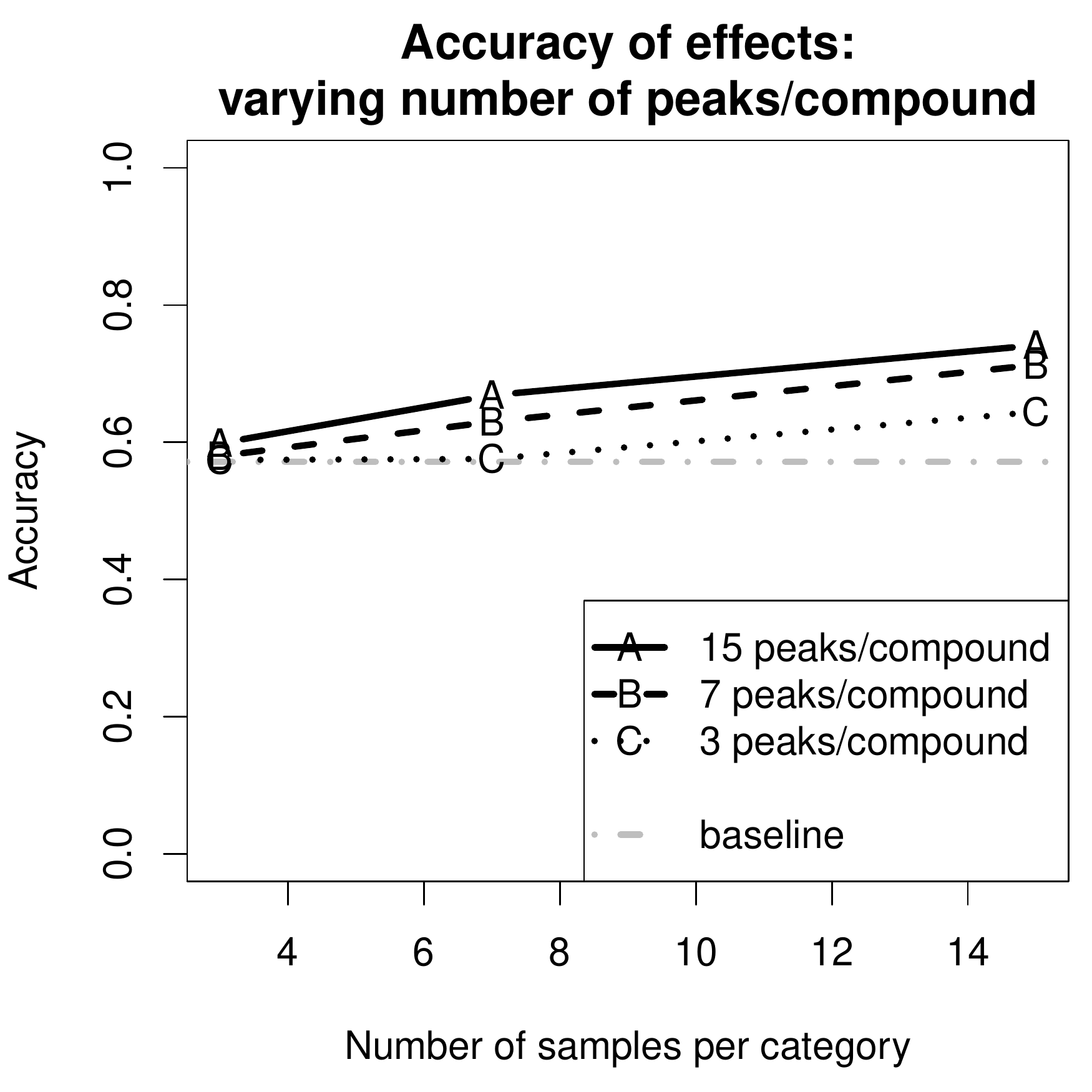}} 
 \caption{The performance of Model~1 improves when more peaks per compound are available in the~simulated data. Curves~A,~B and~C show the~accuracy as a~function of sample-size for simulated data with~15,~7 and~3 peaks per compound, respectively.}
 \label{fig:simExp-Npeaks}
\end{figure}

The performance of the~multi-peak approaches clearly improves, when more peaks from a~cluster are present in the~data~(Figure~\ref{fig:simExp-Npeaks}). This is pronounced at a~high level of noise, when the~observation of a~single peak is unreliable for inferring the~covariate effects. In~a~similar way as in averaging over samples, the~model is able to overcome peak-specific noise also by averaging over multiple peaks.

\subsection{Benchmark data with known changes\\in concentrations}

\subsubsection{Introduction} In the~first demonstration on real UPLC-MS data~\cite{Franceschi11}, we show that Model~1 can infer the~artificial perturbations in a~spike-in experiment more accurately than the~single-peak approach. The~recently-published benchmark data set of apple samples includes a~set of annotated spike-in compounds with increases of~20,~40 or~100~\% in concentrations.

We start with the~raw spectral data~\footnote{The~raw spike-in data~\cite{Franceschi11} is available at~\href{http://cri.fmach.eu/Research/Computational-Biology/Biostatistics-and-Data-Management/download/data/Spiked-Apple-Data}{http://cri.fmach.eu/Research/Computational-Biology/Biostatistics-and-Data-Management/download/data/Spiked-Apple-Data} [Accessed~11.06.2013].} in order to acquire the~shapes of the~peaks in addition to their heights. The~mass spectra are pre-processed using MZmine~2~\cite{Pluskal10}~(Section~3 in Supplementary material).

We evaluate the~approaches by the~MSE between inferred and true covariate effects. If the~cluster contains multiple annotated peaks, the~effect of each annotated peak is evaluated separately for the~single-peak approach. Clusters with no annotated peaks are considered to have a~0~\% true effect and the~effect of the~single-peak approach is inferred based on the~strongest peak of the~cluster.

\subsubsection{Result}All clusters with annotated peaks are specific to one compound. In~the~negative ion mode, all peaks from one compound are clustered together. In~the~positive ion mode, peaks from two compounds are distributed to two and four clusters. Model~1 has a~lower error than Model~3 at all magnitudes of the~true effect with the~strongest improvement occurring at the~small magnitudes~(Table~\ref{table:relativeChange}).

\begin{table}[h!]
\caption{Model~1 yields a~more accurate quantification of the~covariate effects for the~spike-in compounds as well as for the~unchanged non-annotated compounds in the~benchmark experiment. Root mean squared error~(RMSE) between the~inferred and true covariate effects is smaller for Model~1 than for the~single-peak approach~(Single) at all magnitudes of the~true effect~(rows), in both the~positive~(a) and negative~(b) ion modes. Statistical significance of the~differences between errors is evaluated using the~one-sided paired $t$-test on the~null-hypothesis of equal MSEs. Significant differences at confidence levels~95~\% and~99~\% are highlighted by symbols~''*`` and~''**,`` respectively.\label{table:relativeChange}}
\vspace{5pt}
\centerline{
\begin{tabular}{r|cc|c}
\hline
 True covariate effect & \multicolumn{2}{c|}{RMSE} & Corrected $p$-value\\
 & Single & Model~1 & of the~difference\\
\hline
 \multicolumn{4}{l}{a) Positive ion mode}\\
 0~\% & 0.37 & {\bf 0.27} & $<2.2 \times 10^{-16**}$ \\
 20~\% & 0.38 & {\bf 0.19} & $7.3 \times 10^{-4**}$ \\
 40~\% & 0.41 & {\bf 0.27} & $1.2 \times 10^{-2*}$ \\
 100~\% & 0.95 & {\bf 0.82} & $6.6 \times 10^{-2}$ \\
 \multicolumn{4}{l}{b) Negative ion mode}\\
 0~\% & 0.38 & {\bf 0.28} & $<2.2 \times 10^{-16**}$ \\
 20~\% & 0.40 & {\bf 0.18} & $9.2 \times 10^{-4**}$ \\
 40~\% & 0.52 & {\bf 0.27} & $3.2 \times 10^{-4**}$ \\
 100~\% & 0.77 & {\bf 0.60} & $5.0 \times 10^{-2}$ \\
\hline
\end{tabular}
}
\end{table}

Based on the~performance of the~single-peak approach, it is evident that the~signal-to-noise ratio is low even in this type of a~spike-in data set with a~highly controlled experimental setup. Model~1 can significantly improve accuracy at the~quantification of covariate effects hidden under the~noise, particularly when true effects are weak.

\subsection{Lipidomic data from a~gene silencing study}

\subsubsection{Introduction} In the~second demonstration on real UPLC-MS data~\cite{Hilvo11}, we show that Model~1 can infer more consistent covariate effects even though the~actual true effects are unknown. The~data come from a~recent experiment to study the~effects of gene silencing treatments on lipidomic profiles and growth of breast cancer tissue.

Driven by the~lack of ground truth about the~covariate effects, we evaluate the~inferred effects indirectly in two ways: by quantifying the~consistency of the~effects within a~lipid family and by quantifying the~robustness of the~magnitudes of the~inferred effects across the~lipidome in presence of additional noise. Additionally, we investigate the~stability of the~inferred clustering on the~data and qualitatively analyze differences between the~covariate effects of single peaks and the~effects inferred on clusters of peaks by Model~1.

The data include~32 lipidomic profiles of breast cancer cells from the~ZR-75-1 cell line. We infer the~effects of seven distinct silencing interventions on metabolism-\ regulating genes~(Section~4 in Supplementary material) at two time points. The~raw spectra are pre-processed with MZmine~2 as described in the~original publication~\cite{Hilvo11}, in addition to which the~shape similarity of the~peaks is computed.

\subsubsection{Consistency of effect signs}

In the~first task, we quantify the~consistency as the~accuracy at predicting the~covariate effect of a~test lipid given the~model on the~covariate effects of other lipids of the~same family. We examine signs of effects instead of absolute effects because even within a~family of lipids, the~changes have a~high variance and thus cannot be reliably predicted without imposing additional information about the~biological system.

We predict the~signs of the~covariate effects for test lipids in a~three-fold cross-validation setting with 100 randomizations. The~examined lipids include the~annotated members from the~three most abundant families of lipids that have two or more peaks identified with the~clustering model~(Section~4 in Supplementary material). Further, we study the~influence of noise to the~consistency by adding independent normally distributed noise to the~peak intensity observations.

\begin{figure}[h!]
 \centerline{\includegraphics[width=0.65\textwidth]{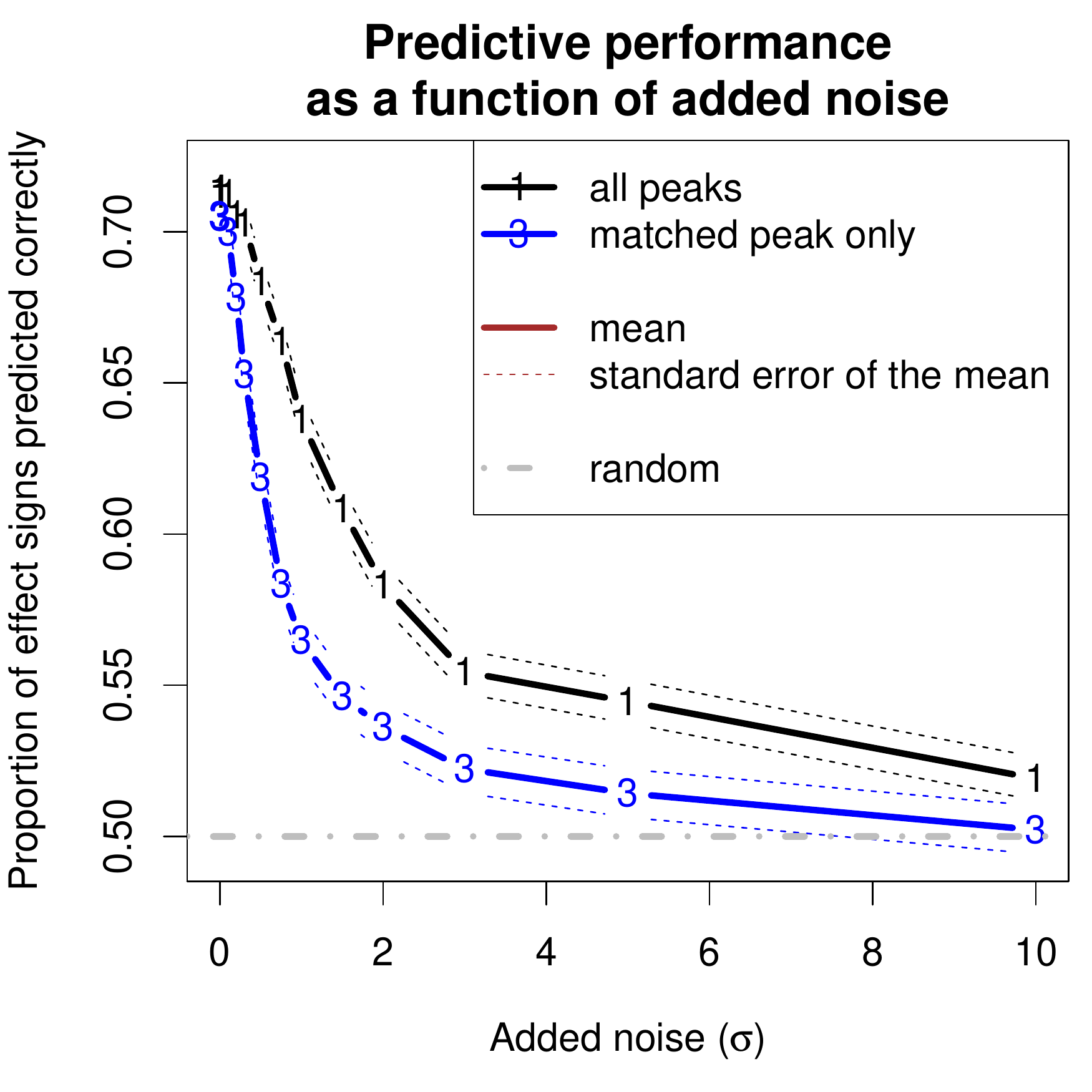}} 
 \caption{Model~1 yields a~better accuracy at the~prediction of signs of covariate effects for previously unseen lipids in the~lipidomic gene-silencing data set. The~difference to Model~3 becomes pronounced when simulated noise is added to the~data. The~prediction is based on the~inferred covariate effects of compounds from the~same lipid family and is done in a~cross-validation setting. In~the~task, the~effects of the~seven gene-silencing treatments are predicted on the~three most abundant families of lipids in two time points. Points~$\sigma=0$ and~$\sigma>0$ on the~x-axis show the~prediction accuracy (y-axis) for the~original data and the~data with added noise, respectively.}
 \label{fig:familiesCV}
\end{figure}

Model~1 proves more consistent than Model~3 at all noise levels~(Figure~\ref{fig:familiesCV}). When no noise is added and also at moderate levels of noise, both approaches perform clearly better than expected by random chance. When noise is added, Model~3 suffers more and its performance falls to the~random level more rapidly. If our assumption about the~general similarity of lipids within a~family is true, we can conclude that Model~1 yields infers the~covariate effects more consistently.

\subsubsection{Robustness of effect magnitudes}

To evaluate the~inferred effects at the~scale of the~entire observed lipidome, we examine the~consistency of inferred covariate effects between the~original and noise-added data sets. This experiment simulates the~situation where the~true effects are known~(effects from the~original data set), but the~data based on which the~effects are inferred are noisy~(the added-noise data set). To measure the~consistency, we compute the~Spearman correlation between the~covariate effects inferred from the~original and the~added-noise data sets for all clusters with two or more peaks.

\begin{figure}[h!]
 \centerline{\includegraphics[width=0.65\textwidth]{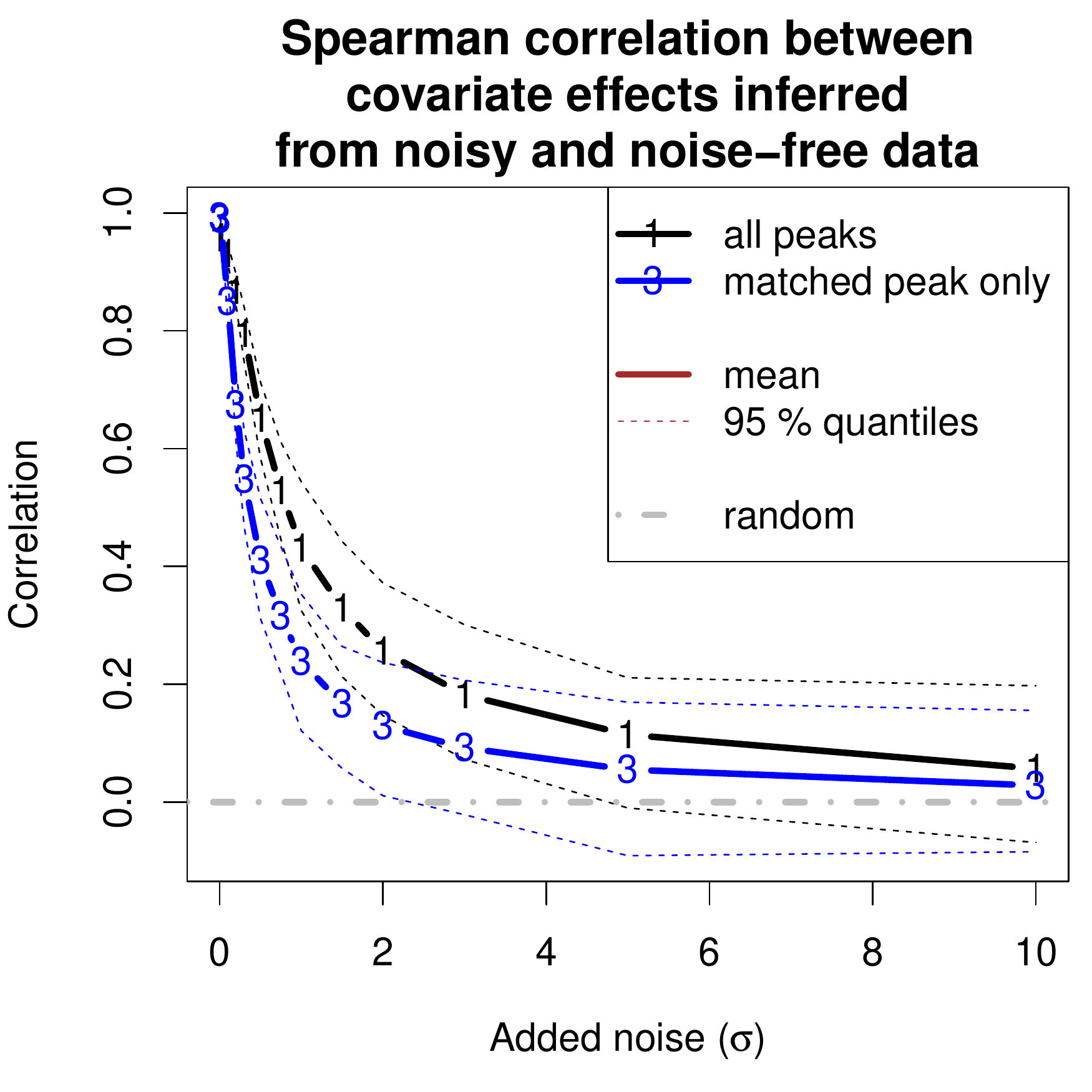}} 
 \caption{The covariate effects inferred by Model~1 are more robust to noise compared to Model~3. At moderate levels of noise, which is the~regime of many biological experiments, the~confidence intervals over~100 randomizations do not overlap at all. The~robustness is quantified as the~Spearman correlation~\mbox{(y-axis)} between the~effects inferred from the~noisy and noise-free versions of the~lipidomic gene silencing data set as a~function of the~level of noise~(x-axis).}
 \label{fig:effectRankCor}
\end{figure}

Model~1 proves more consistent than Model~3 at all noise levels~(Figure~\ref{fig:effectRankCor}). The~confidence intervals from the~100 randomizations do not overlap at all at moderate levels of noise. This gives indication that for a~typical biological data set, there is clear benefit from the~analysis of multiple peaks, leading to reduced variance of the~result.

\subsubsection{Stability}

As the~proposed approach is sensitive to the~inferred clustering of the~data, we analyze the~stability of the~inferred clustering on biological data, using the~lipidomic gene silencing data as a~case. We test the~influence of the~concentration parameter~$\alpha_\text{DP}$ in the~Dirichlet process clustering model. The~clustering result for the~lipidomic gene silencing data turns out to be very robust to changes in the~magnitude of the~concentration parameter~(Figure~2 in Supplementary material). As~expected, the~number of clusters increases as the~preset value of the~concentration parameter increases but the~relative change is small. Also the~change in the~consistency of the~learned covariate effects is small.

\subsubsection{Qualitative analysis}

Finally, we give concrete examples of potential findings that the~approaches can uncover and demonstrate how analysis based on a~single peak may lead to a~different outcome depending on the~choice of the~peak.

\begin{figure}[h!]
 \centerline{\includegraphics[width=0.9\textwidth]{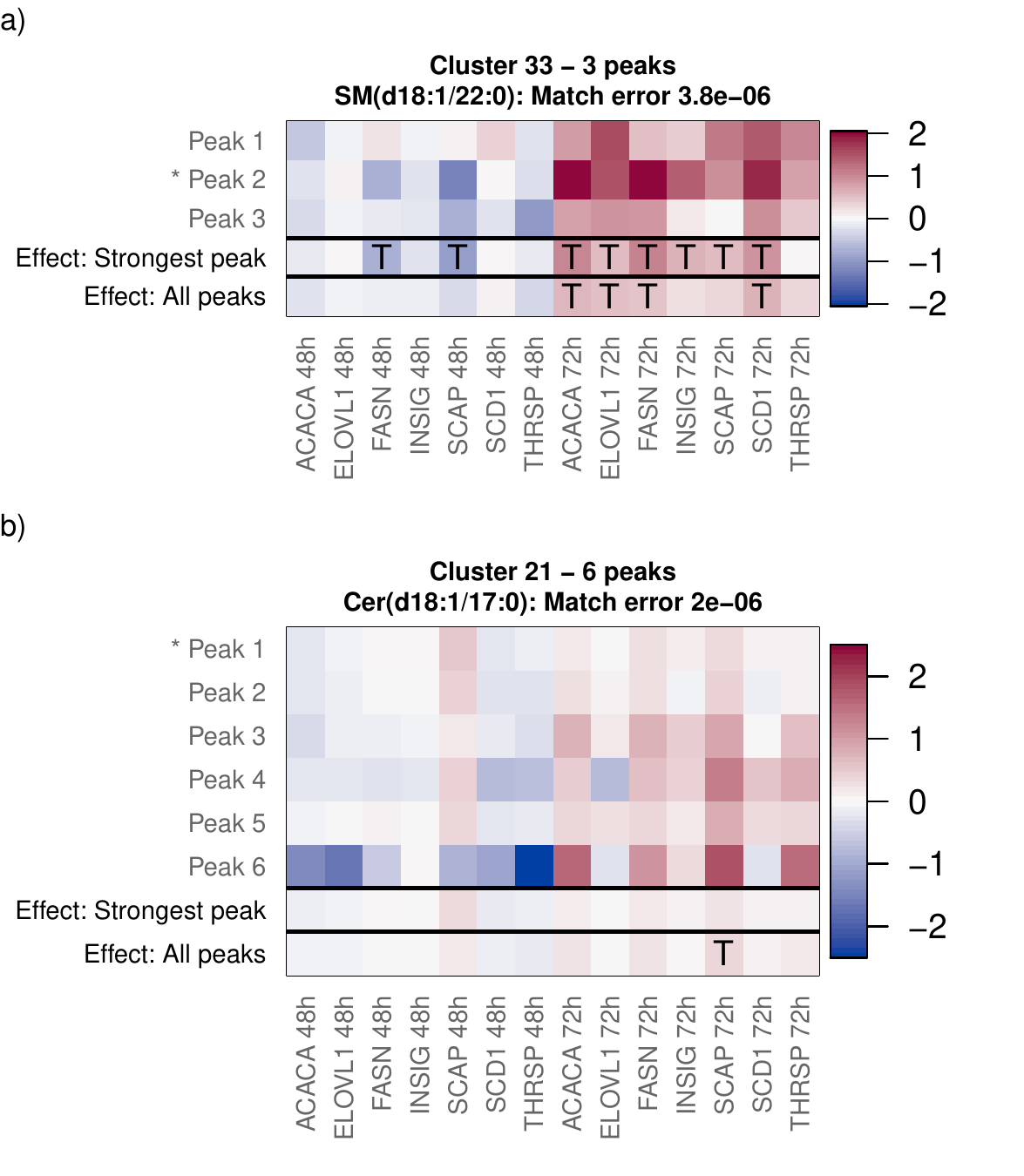}} 
 \caption{Example clusters of peaks from the~lipidomic gene silencing data with differences in the~covariate effects inferred based on a~single peak and multiple peaks. There is noticeable variation between peaks associated with one compound. The~choice of the~representative peak thus affects the~inferred covariate effects. The~heat maps show changes in the~lipid concentrations driven by the~gene silencing interventions~(columns). Covariate effects inferred by Models~3 and~1 using a~single peak and all peaks, respectively, are shown on the~two bottom rows of each heat map. The~$\log_2$ fold changes of each peak associated with the~compound are shown on the~top rows. Changes that by the~magnitude fall to the~top-5~\% across the~entire observed lipidome are highlighted by the~symbol~''T.`` Top~(a): The~sphingomyelin~\mbox{SM(d18:1-22:0)} with three peaks. Many strong changes for~\mbox{SM(d18:1-22:0)} become weaker when they are inferred based on all three peaks. Bottom~(b): The~ceramide \mbox{Cer(d18:1-17:0)} 
with six 
peaks. The~effect of the~SCAP silencing for \mbox{Cer(d18:1-17:0)} at~72~hours becomes strong when it is inferred based on all six peaks.}
 \label{fig:vttExp-lipidHeatPlot}
\end{figure}

The intervention-driven changes of individual peaks from two lipids along with the~covariate effects inferred by Models~1 and~3 are shown in Figure~\ref{fig:vttExp-lipidHeatPlot}. In~the~case of the~sphingomyelin~SM(d18:1/22:0), there are strong covariate effects inferred by Model~3 but many of these effects become weaker when inferred based on multiple peaks by Model~1. On~the~contrary, Model~3 infers weak covariate effects for the~ceramide~Cer(d18:1/17:0) but based on multiple peaks and Model~1, one of the~effects is actually among the~top-5~\% strongest effects across the~observed lipidome. These examples underline the~fact that conclusions based on mass spectral data are highly sensitive to the~choice of peak, but the~variance therein can be reduced by model-based integration of data from multiple associated peaks.

\section{Conclusion}

We have empirically demonstrated that a~model-based integration of multiple peaks can lead to an~improved accuracy in the~inference of covariate effects, and introduced a~novel method for this task. While the~sample-size is always restricted by external constraints such as the~experiment budget or the~availability of suitable patients, the~inference based on multiple peaks gives a~shortcut to extracting more information from the~limited set of samples, thereby directly addressing the~``small~$n$, large~$p$'' problem. However, some types of systematic measurement error, such as some batch effects, escape this treatment and can only be reduced by introducing independent replicates. In~this work, we have shown that additional peaks are especially useful when the~signal-to-noise ratio is low and the~differences between sample groups are small.

Mass spectral data are noisy to the~extent that analysis of small sample-size data sets based on individual spectral peaks is unreliable. The~result of this work indicates that additional peaks are useful to the~analysis by increasing the accuracy of the~inferred covariate effects and by reducing the~variance in the~estimates. Moreover, the~result highlights the~risk of relying on a~single peak in the~comparative analysis of profiles and especially in biomarker discovery. As~demonstrated, the~inferred covariate effects may change drastically depending on the~peak used, when relying on a~single peak in the~analysis.

We suggest that all the~detected peaks that can be associated with a~compound should be taken into account in the~comparative analysis. This is possible through the~two-step generative modeling approach presented in this work: (1)~the~identification of peaks that can be associated with one compound by clustering the~peaks based on their shape similarity and (2)~the~inference of covariate effects on the~clusters, each representing one compound.

\section*{Acknowledgements}

The~authors would like to thank Sandra Castillo, Peddinti~V. Gopalacharyulu, Mika Hilvo and Matej Ore\v{s}i\v{c} for providing data and for useful discussions. The~authors would also like to thank R\'on\'an Daly and Joe Wandy for useful discussions.

This work was supported by the~Academy of Finland~(Finnish Centre of Excellence in Computational Inference Research~COIN,~251170; Computational Modeling of the~Biological Effects of Chemicals,~140057), the~Finnish Foundation for Technology Promotion~(to~TS) and the~Finnish Doctoral Programme in Computational Sciences~FICS~(to~TS). 

The~calculations presented in the~work were performed using computer resources within the~Aalto University School of Science "Science-IT" project.

\bibliographystyle{abbrv}
\bibliography{peakANOVA-Suvitaival}

\end{document}